# Integrated Science, Technology, Engineering and Mathematics (STEM) Education through Active Experience of Designing Technical Toys in Vietnamese Schools

Le Xuan Quang[1], Le Huy Hoang[1], Vu Dinh Chuan[2], Nguyen Hoai Nam[1*], Nguyen Thi Tu Anh[1] and Vu Thi Hong Nhung[1]

[1]*Faculty of Technology Education, Hanoi National University of Education, Vietnam.*
[2]*Ministry of Education, Vietnam.*

***Authors' contributions***

This work was carried out in collaboration between all authors. Author LXQ contributed to the design of the research project, and drafted the manuscript. Authors NTTA and VTHN designed and made technical toys; authors LHH and VDC contributed to the content of the draft manuscript. Author NHN contributed to the framing and final writing of the article. All authors read and approved the final manuscript.

## ABSTRACT

**Aim:** STEM has attracted great consideration. The purpose of research is: (i) study STEM education; (ii) explore STEM education with the creative and experiential activity; (iii) suggest applying STEM education by designing technical toys for the middle school student.
**Study Design:** This study used a qualitative approach to carry out teaching integration for STEM education.
**Place and Duration of the Study:** The study applied to teaching the technological field in Vietnamese middle schools. The design performed at the Faculty of Technology Education, Hanoi

---

*Corresponding author: E-mail: namnh@hnue.edu.vn;*

National University of Education (HNUE), Vietnam in April 2015.
**Methods:** This study used the integrated approach to design subjects for STEM education.
**Results:** Two procedures for integration undertook with analysis. A sample of producing technical toy was consistent with developing students' competencies.
**Conclusion:** Integrated approach to STEM education through designing technical toys is possible. Recently, there has been a booming interest in Integrated Science, Technology, Engineering and Mathematics (STEM) education [1,2], but the approaches to STEM still remains controversial in diverse educational contexts. This study addressed this issue by exploring STEM education with the use of creative and experiential activities in a Vietnamese educational context. It also proposed a practical model for integrating STEM into teaching technology in secondary schools by designing technical toys. The implementation of the practical model suggests the possibility in using the integrated approach to STEM education through designing technical toys for middle school students in Vietnam. By applying the subject knowledge domains to solve real world problems and settings, the students can experience the benefits of a concrete and active learning in a meaningful and practical context. The multidisciplinary and interdisciplinary integration approaches are consistent with the development of the students' competencies.

## 1. INTRODUCTION

The integration of Science, Technology, Engineering and Mathematics (STEM) has been an area of interest since it was first addressed in education in the United States in the early 1990s. This integration was considered a solution to the educational reforms in the United States when the society needed to provide highly qualified labors with complex technology and engineering skills to perform in the high-tech knowledge-based economy. Indeed, the primary focus of STEM education is to prepare competencies in these multi-disciplines for students in order to meet the requirements for the 21$^{st}$ century workforce [3]. As William [2] stated, the shortage of STEM workers threatening the United States competitiveness has urged the STEM educational reforms in this country. It then has rapidly spread across many countries. National reports on STEM educational reforms has been published in several countries as Australia, England, Scotland, the United States [4-7]. The K-12 STEM curriculum in each educational context has been designed with STEM subjects from integrative cross-disciplinary approaches. It creates a connection between learners' STEM educational experiences and their future careers [6,8-12]. Moreover, each STEM curriculum has focused on the development of interests in STEM careers for young learners through extra-curricular activities, contests [8,13]. However, the approaches to STEM are different in countries because of the political, social and technological history [2].

The efficacy of STEM education has been addressed in the literature over the last decades. Only within four years, from 2007 to 2010, over 1,100 articles discussed this integration in education [1]. It helps students not only develop their skills but also construct their awareness of science and engineering concepts through experiential learning methods [14-16]. The interdisciplinary promotes the middle and early secondary school student learning and STEM activities [17]. Although the literature focusing on the methods of STEM integration, efficacy of this process, for example, has been studying and up-to-date [1], the caution about curriculum, clarity, vocational and general education, alignment, epistemology, and goals still need to be explored in a variety of educational contexts [2]

In Vietnam, although the education authorities and scholars have recently paid attention to STEM [18], controversial issues have still been under consideration. In the curriculum, the STEM integration has been applying to general subjects in the primary education. It has not been implemented in the middle and high school levels, even though it should be noted that middle school is an important phase for students because it establishes the foundation for and orientation to their careers. School teachers, moreover, principally have provided a focus on the content knowledge rather than the students' practical competencies. The shortage of practical experience in learning causes some obstacles for Vietnamese students. Vietnamese students have been characterized as passive learners who have difficulties in problem solving in reality, creative thinking and who are deficient in technical skills such as information and communication technology (ICT), consequently. Thus, the inadequate preparation for students as the future qualified workforce is obvious in this educational context.

Many studies showed the correlation between student learning and educational teaching aids including educational toys. Educational toys can be various types as handmade, machine-made or robotics. Having studied Hong Kong students' attitudes toward technology (PATT) in the middle school, Volk [19] found that the students showed positive learning attitudes when they studied with technical toys. This finding has aligned with other studies regardless the different geographies and students' ages [20,21]. Learning is actively constructed when students take part in toy design [22,23]. The students' learning experience with such toys brings about the efficiency [24,25].

Thus, this study aims to explore the application of STEM education in a situated learning context in Vietnam. It focuses the efficacy of STEM education when the pre-service students from a pedagogical university in Vietnam design technical toys for the students at the middle-school level. This study also investigates how STEM education works with this creative and experiential activity in this context.

## 2. MATERIALS AND METHODS

## 2.1 Understanding STEM Education

Diverse definitions of STEM education has been discussed in the literature from different approaches such as a silo, embedded and integrated one.

### 2.1.1 The silo approach

In the silo approach, teachers coach individual STEM subject separately [26]. In each subject, they focus on the core knowledge. Students are, therefore, expected to gain deep understanding of the course contents. In this learning process, teachers play an important role. They attempt to impart the knowledge through their high-standard instructions for their students. However, students learn to know the knowledge but not experience and gain the learning knowelldge by doing [27].

This approach has some drawbacks. As Dickstein [28] supposes, it prevents the contribution of STEM when students are passive in learning. It is likely that students misunderstand the integration, which naturally occurs among STEM subjects in the real world [29]. Students also feel little motivation because teachers mainly impart the knowledge through the lecture-based method rather than a hands-on approach as they wish [28].

### 2.1.2 The embedded approach

The embedded approach highlights the real world and problem-solving techniques within social, cultural, and contexts of knowledge domains [30]. The instruction tends to be more effective because it enables the students to reinforce what they learned in other classes [31]. Contrary to the silo approach, this embedded approach encourages learning through various contexts [32].

However, the embedded approach has weaknesses concerning the missing design in evaluation and assessment [30], and the learning fragmentation [33]. A learning disruption may occur if the teacher-student interaction for constructing the embedded knowledge is interrupted for correcting feedback [34]. Students cannot associate the set-in contents of the lesson and, as a result, tend to lose the whole lesson.

### 2.1.3 The integrated approach

In the integrated approach, the STEM content areas are mixed and learnt as one subject [27,35]. Students are expected to use multidisciplinary STEM concepts to solve real world problems [36]. This learning process appears to activate the concentration and to increase the motivation in STEM content areas, especially with young learners [37,38].

According to Wang [36], there are two kinds of integrative instructions called multidisciplinary and interdisciplinary ones. While the multidisciplinary instruction develops students' abilities to connect the knowledge domains among specific subjects, the interdisciplinary one incorporates these knowledge domains and individual subject skills. Therefore, the interdisciplinary method asks students to incorporate cross-subject contents with critical thinking, problem-solving skills, and knowledge to solve a real world problem. Nevertheless, the multidisciplinary one tends to create a stronger connection among a variety of subjects in numerous classrooms at a different time as faculties' corroboration.

It is likely that the best approach to STEM instruction is the integration in accordance with many above studies. However, the integration of STEM subjects may be detracted from the integrity of any individual STEM subject because of the disparities among the underlying epistemological assumptions of STEM disciplines [2]. Besides, the lack of the general structure of a lesson may limit students' comprehension, known as potpourri effect [39]. In this case, teachers probably fail to create one common objective despite the material incorporation from each discipline.

Each approach has strengths and weaknesses that need further investigation. Teachers should evaluate the subject knowledge domains and choose the best teaching approach [40].

### 2.2 STEM Education with the Creative and Experiential Activity

Torrance [41] defined the creativity as the process of sensing gaps or disturbing, missing elements; forming ideas or hypotheses concerning them; testing these hypotheses; and communicating the results, possibly modifying and retesting the hypotheses. To promote the creativity in science classrooms, Dass [42] discussed some strategies such as visualisation, divergent thinking, open-ended questioning, consideration of alternative viewpoints, generation of unusual ideas and metaphors, novelty, solving problems and puzzles, designing devices and machines, and multiple modes of communicating results.

Studies show the creativity, problem-solving and design as essential skills in students' STEM development [43,44]. While science is a process of investigation and inquiry, engineering is a process of design that requires a blend of knowledge and creativity [45]. Those skills possibly develop well with hands-on experiential activities especially when these activities create an environment of active learning.

Active learning is an instructional method that engages students in the learning process to promote the learning outcomes [46]. Through the created learning environment, knowledge is directly experienced, constructed, acted on, tested, or revised by learners [47] and the interaction between stakeholders (for example: Teacher, students, materials...) is improved. Students may develop many required skills through this active learning process as communication, higher-level thinking, collaboration, problem-solve, creativeness, for example, with a positive attitude and motivation as well. Problem-based, experiential, and inquiry-based learning are commonly known as forms of active learning. However, the formal and structured methods need designing with a strong basis or background within any given area for students at the early stage, then the effectiveness can be strengthened by the active learning [48].

With the activities as extra-curricular, contests, for instance, the students' interest and motivation in STEM careers improve [8,13]. In fact, the annual creative, experiential contest held by Vietnamese Ministry of Education draws much attention and attendance of students, as well stakeholders (parents, teachers, companies, society, for example). Students experience activities as designing, studying and producing many materials applied to learning, working... Therefore, they appear to consolidate the motivation and understand the meaning of an experiential activity and the supplementary concern in studying STEM.

### 2.3 STEM Education with the Technical Toy Design

Studies have shown the benefits of educational toys in teaching. In Sirinterlikci's study [22] of the challenges for students between grades 5 and 8 of using educational toys and its application, he argues that hands-on learning as a means of promoting the interest in science for young learners. Students had positive attitudes toward engineering-related knowledge. By doing that, students explored a set of ideas and used high-level thinking in deciding and solving problems, repeating the process of a scientific inquiry used by experts in STEM fields. Having applied knowledge to the real world problems and settings, students could experience learning in a practical and meaningful way [49]. These findings are consistent with other work about African-American $4^{th}$ -$6^{th}$ grade girl students' attitude toward STEM by activities including toy design [23]. The students' enthusiasm for learning, their confidence and ability in science as well as their interest in STEM careers, therefore, increased.

Designing technical toys is applicable to secondary students regarding their psychology and abilities because it helps them familiarize with the features of an engineering design. However, what may concern teachers is the students' differences in the ways of thought and action. According to NAGB [50], engineering design requires a systematic and creative approach for addressing a challenge that students may encounter in a problem solving. However, school students tend to stick to the first solution that comes to their mind to solve it. In order to train this practice, it is noted that, students need to think carefully over the procedure for problem solving such as defining the problem, making several solutions, and testing, evaluating, revising and testing again during the engineering design before they end up with a solution.

Design and exploiting the top, Worch [51] agued that the students' competencies in maths and physics developed. He suggested three phases, namely engaging, exploring and extending, to utilise the top for the STEM education. The materials for the top were inexpensive and easily obtainable at any discount or craft store. Thus, making a technical toy with cheap materials is possible.

## 2.4 Procedure for Technical Toys DESIGN with STEM Education

The integrated approach is radical in the primary school but a real problem in the secondary school. Firstly, the school curriculum, as William [2] pointed out, hardly changes. Secondly, a teacher teaches all subjects at the primary school level, but each teacher is in charge of one subject in the secondary. Currently, it is not practical to combine all the STEM subjects into a whole for an individual teacher in secondary schools. Therefore, teachers should find out the intersection of subjects to integrate through their collaboration. Teachers can choose an embedded or an integrated approach to carrying out STEM sections. This study employs the integrated approach.

The proposed procedure was designed and divided into 5 steps for teachers as follows:

Step 1: Teachers must study subjects and contents of subjects related to STEM (Technology, Maths, Physics, etc.).
Step 2: Teachers identify the intersection between content subjects and learning outputs (e.g. knowledge, skills and attitude for students) relating to STEM and evaluate the possibility of integration. They study learning material of each subject, including textbooks and others. This can be done in both formal and informal ways, for example, from the Internet to concrete an intersection subject for integration.
Step 3: Teachers decide types of technical toys that may include most of the knowledge in STEM subjects that they want their students to study. Questions should be designed to assist student in finding the connection between content knowledge and technical toys.
Step 4: Teachers design technical toys and evaluate the possible application of the toys for STEM education. After testing and judging, teachers may modify or redesign these toys to increase benefits of STEM education.
Step 5: Teachers organise a classroom to instruct students to make a technical toy. This is in a form of a creative and experiential activity. Many teaching methods can be used; for example, problem-based, project-based, inquiry-based and so forth.

The procedure for teachers to follow is briefly described in Fig. 1 bellows:

### 2.4.1 The needs for technical toy design

To optimise technical toys, the design must meet the following needs:

- Multi-functions and multi-choices: The multi-functions of the technical toy are a possibility to work several demands. The multi-choices are a flexibility in designing, producing, for example, as a detail depicted in several ways; a product built with different procedures.
- Integration: technical toys related to various subjects, but it is presented as a whole. The content knowledge of technical toys is associated with all selected subjects such as science, maths, technology, etc.

With features, teachers analyse the scientific basis of design and the applicable possibility.

A technical toy must satisfy other specifications as follows:

- Educational features: Matching curriculum and being consistent with students' abilities of learning.
- Economical features: Being produced from common materials which are simple, easy to find and being made with simple tools
- Technical features: Meeting essential strength and accuracy.
- Safe features: Conforming to students psychology and health.

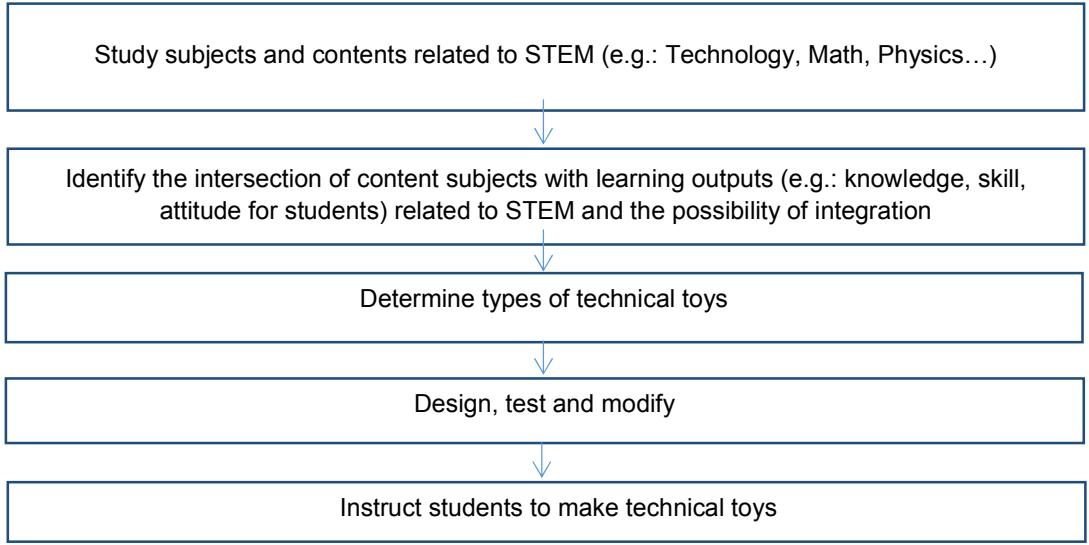

**Fig. 1. Procedure for technical toy's design for teachers**

In the classroom, teachers guide students to the procedure as follows:

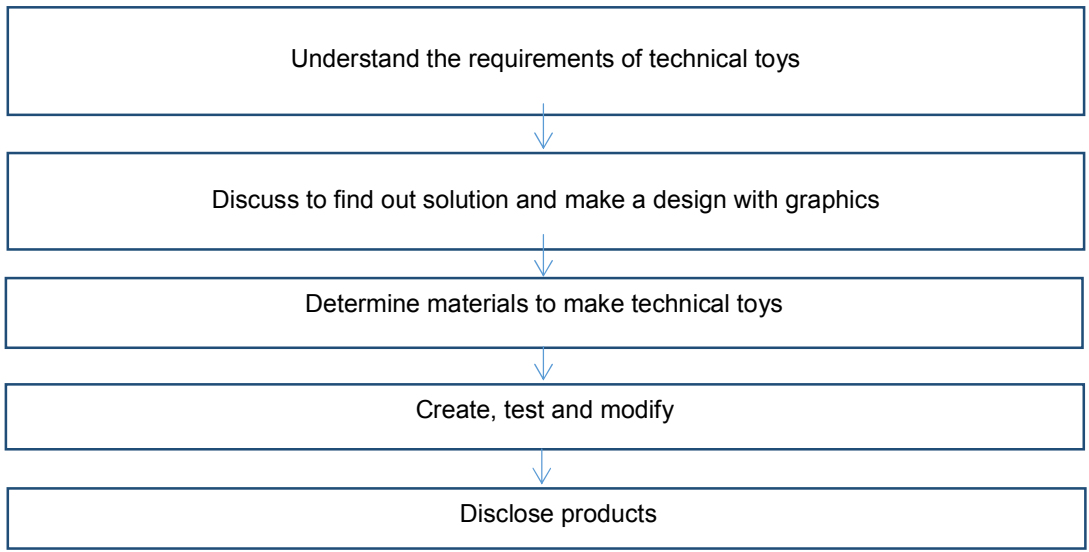

**Fig. 2. Procedure for technical toy's design for students**

Step 1: Students must understand the needs of the technical toys they will make (for example: Tasks, features, styles...). They may watch the samples made by the teacher as a suggestion.

Step 1: Students discuss in a group to find out the solution: what they like best and what is the best. They make a model design with graphics. To have a good design, students actively collaborate as they apply learned knowledge, imagine themselves or search some suggested information from other as textbooks, the Internet, or recommendations of the teacher.

Step 3: Students settle materials to make technical toys. They choose suitable materials for the technical toys and tools to produce.

Step 4: Students create technical toys with the design and materials. They test and modify the product if it meets the requirements or not. In this phase, students experience hands-on activities, have opportunities to practise and perform. They can apply the knowledge have learned from the previous lessons and their social experiences. The teacher should encourage students to devote themselves in any idea that makes them excited.

Step 5: Students present the designed products in class. They may feel proud of their products, and be interested in STEM education.

## 3. RESULTS AND DISCUSSION

Based on the procedure of technical toy's design, the authors offer the experiential activity for the $8^{th}$ grade student in designing a mini-racing car.

According to the Vietnamese curriculum, the needs for students at $8^{th}$ grade to integrate in activities of designing a mini-racing car are shown in the Table 1.

The teacher introduces the sample of mini-racing car with the task of running on and going over the ramp with minimum passing time.

The minimum requirement for construction is attaching the car to the essential model (The car has a backbone chassis, actuator). Students can change or attach the extra details to optimise tasks. They may operate individually and in groups.

Materials are supplied to each group, including 01 plywood, wheels, spindles, mini motors bevel wheels, belts, batteries, switches, screws…; tools include a small saw, scissors, pliers, a screwdriver, glue…

**Table 1. Competencies in designing a technical toy "a mini-racing car"**

|  | Technology 8 | Maths 8 | Physics 8 |
|---|---|---|---|
| Competencies | - Competencies in design: applying engineering drawing skills to design car style<br>- Competencies in technical activities: using simple tools such as pliers, hammer, scissors…; understanding the engineering details and assembly…; applying the transmission knowledge to choose the transmission drives such as a belt drive, gear drive…; applying the wire connector to turn on/off and operating the motor in proper turns | - Competencies in algebra: measuring the long wheelbase, calculating the transmission ratio, the length between details<br>- Competencies in geometry: analysing and cutting the polygons with simple tools, e.g. Compass, calculating area and evaluating properly for the components | - Competencies in mechanics: understanding the kinds of motion, friction; analysing and evaluating the factors affected by the speed of the car, e.g.: Friction, weight of the car<br>- Competencies in electrics: understanding the transmission energy from the electrical energy to the mechanical energy, evaluating the engineering power |

The multidisciplinary and interdisciplinary integration approaches can be determined as follows:

**Table 2. The integrated approach to design a technical toy "a mini-racing car"**

| Competencies from subjects | Figure |
|---|---|
| Applying knowledge of the technical drawing skills and the technological subjects to design backbone chassis, including size, styles. | 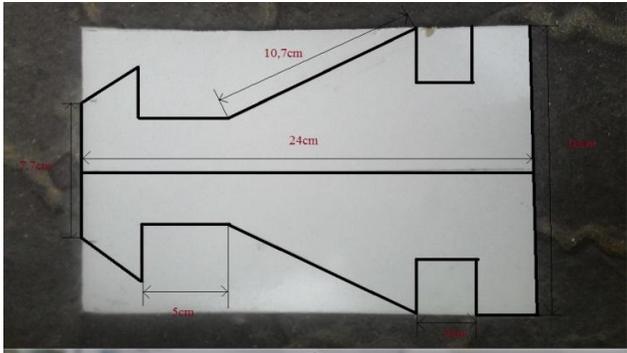 |
| Applying knowledge of the geometry from the maths subjects to cut polygon and calculate the area of chassis. Students use the simple tools as compass… and evaluate properly parts. | 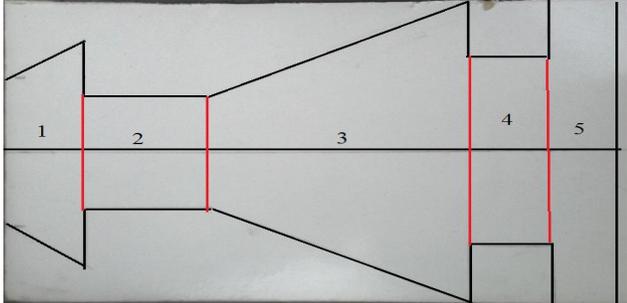 |
| Applying knowledge of turning from the technological subjects to produce the details. Students learn how to use tools and revise the understanding from the previous lessons to complete in practice. | 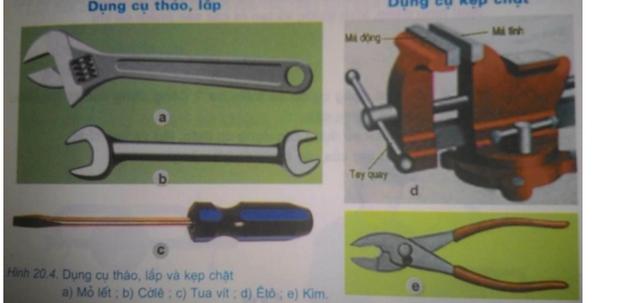 |
| Applying the understanding of the machine details and attachment from technological subjects to connect the components. By revising and understanding learning information, students carry out experientially. | 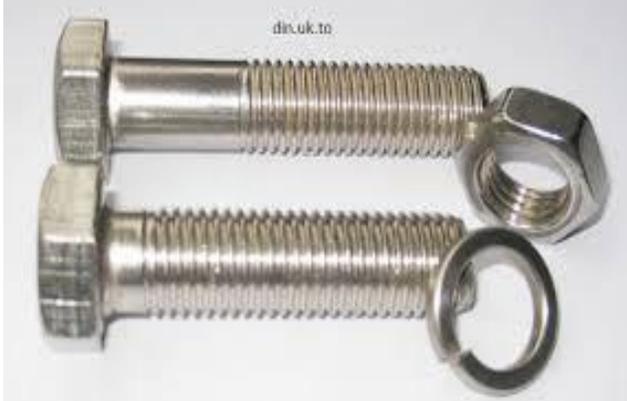 |

| Competencies from subjects | Figure |
|---|---|
| | 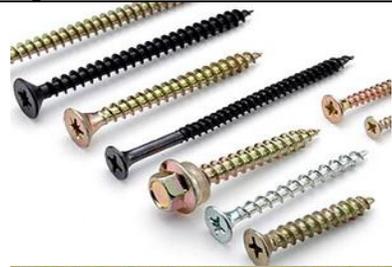 |
| Applying knowledge of transmission and transformation, the motion of the technological subjects to choose transmission instance, for example: sing a belt transmission. | 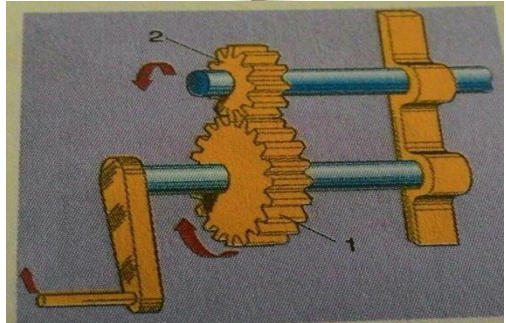 |
| Applying knowledge of the technological subjects to calculate speed and evaluate instances to speed up. Students also need the knowledge that they have learnt in physics lesson to calculate the power and the force. They estimate the reasons affected by the speed of the car. | 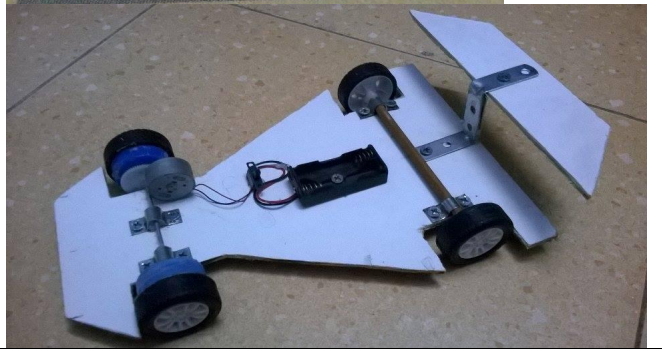 |

While students do the project, the teacher plays a role as an instructor. He or she needs to manage the groups in class, to elicit some questions to promote critical thinking in each group work. Students should be encouraged to exchange the ideas and to improve the sample design. The teacher can use various tactics to have the best achievements [42,52,53]. The concern is that students constrained a strong basis or background within any given area for the experiential to strengthen the learning effectiveness [48].

With the framework, the integrated approach can be undertaken for other technical toys, rowing robot, mini-fan, lift, for example. With such toys, students can work out several possible solutions, for example, the power for a car operation from battery or the wind. The materials for such toys are from second-hand or low cost, obtainable and familiar as another study of hands-on product [24]. They must be safe for the students' health. Exploiting the technical toys in the proper context could develop the STEM skills as the proof of other works [24,25,51].

## 4. CONCLUSION

In conclusion, the integrated approach for STEM education with technical toy design is possible because it allows the application of subject knowledge domains to the real world problems and settings. Students, thus, can experience the benefits of a concrete and effective learning. Students not only study the static knowledge in school but also experience its dynamic learning. The multidisciplinary and interdisciplinary integration approaches with subjects are consistent with the development of students' competencies listed by the Vietnamese Ministry of Education. However, the achievement will be better if the curriculum changes to promote a strong integration among subjects as a whole, and that needs further investigations.

Acknowledgments: The work reported here was part of the doctoral dissertation of the first author. The authors thank to the support in working condition of Faculty of Technology Education, HNUE, Vietnam.

**COMPETING INTERESTS**

The authors declare that they have no competing interests.

*Peer-review history:*
*The peer review history for this paper can be accessed here:*
*http://sciencedomain.org/review-history/10369*